\documentclass[showpacs,preprint,aps,draft]{revtex4}
\usepackage{graphicx}
\usepackage{dcolumn}
\usepackage{bm}
\usepackage{latexsym}
\usepackage{amsfonts}
\usepackage{amssymb}
\usepackage{amsmath}

\begin{document}

\preprint{}

\title{ Braneworld Cosmological Perturbation Theory at Low Energy}

\author{Jiro Soda}
\email{jiro@tap.scphys.kyoto-u.ac.jp}
\author{Sugumi Kanno}
\email{sugumi@tap.scphys.kyoto-u.ac.jp}
\affiliation{
 Department of Physics,  Kyoto University, Kyoto 606-8501, Japan
}%

\date{\today}

\begin{abstract}
 Homogeneous cosmology in the braneworld can be studied without solving 
 bulk equations of motion explicitly. The reason is simply because the
 symmetry of the spacetime restricts  possible corrections in the
 4-dimensional effective equations of motion. It would be great if
 we could analyze cosmological perturbations without solving the bulk.
 For this purpose, we combine the geometrical approach and the low energy
 gradient expansion method to derive the 4-dimensional effective action. 
  Given our effective action, the standard procedure to obtain
  the cosmological perturbation theory can be utilized and the temperature
 anisotropy of the cosmic background radiation can be computed
 without solving the bulk  equations of motion explicitly.
\end{abstract}

\maketitle


\section{Introduction}

  It is believed that the initial singularity problem of the cosmology 
  does not arise in the superstring theory. It is known
  that a clear prediction of the superstring theory is existence of
  extra-dimensions. This apparently contradicts our experience.
  Fortunately, the superstring theory itself provides
  a mechanism to hide extra-dimensions, which is the so-called braneworld 
  scenario where the standard matter lives on the brane, 
  while only the gravity can propagate in the bulk space-time. 
   A possible realization of the above scenario is a two-brane model 
   proposed by Randall and Sundrum~\cite{RS1}. 
   Here, we concentrate on this particular model as a playground
   for studying an interplay between the bulk and the brane
   during the cosmological evolution.
 
 So far, no concrete prediction concerning the evolution of cosmological
 perturbations has been made. The difficulty in solving the bulk with
 moving boundary condition causes this situation. 
 Interestingly, as to the background homogeneous universe,
 this difficulty can be circumvented owing to the spacetime symmetry.
 In this simplest case, the effect of the bulk only comes into 
 the effective 4-dimensional effective equations as the dark radiation. 
 The point here is that we need not
 to solve the bulk equations of motion to obtain the information of the bulk.
 Needless to say, if this point can be extended to the inhomogeneous universe,
 the great progress can be expected. What we want show in this paper is that, 
 at low energy, this is indeed the case,  namely, one can know the 
 effect of the bulk geometry on the cosmological evolution of fluctuations
 without solving equations in the bulk. 
 For this purpose, we combine the derivative expansion of the action
  and the geometrical approach to derive the 4-dimensional 
 effective action with  KK effects for the two-brane system.  
  Our new method  gives not only a simple re-derivation 
  of known results~\cite{KS1},  but also  a new result, 
  i.e. the effective action with KK corrections~\cite{Sugumi}. 

The organization of this paper is as follows:
 In sec.II, we review a geometrical method and 
 discuss the homogeneous cosmology.
 In sec.III, we explain the gradient expansion method. 
 In sec.IV, our strategy to obtain the effective action is illustrated.
 In sec.V, we present the KK corrected effective action and discuss
 the implications of our results on the cosmological perturbation theory.
 In the final section, we summarize our results. 


\section{Geometrical Approach and Homogeneous Cosmology}

First, let us review the geometrical approach. 
Without losing the generality, we can take the Gaussian normal coordinate
 system in the vicinity of the brane. Then, the metric can be written
 as
\begin{eqnarray}
ds^2 = dy^2 + g_{\mu\nu} (y, x^\mu ) dx^\mu dx^\nu \ .
\end{eqnarray}
Defining the extrinsic curvature
\begin{eqnarray}
 K_{\mu\nu} = - {1\over 2}{\partial \over \partial y} g_{\mu\nu} \ ,
\end{eqnarray}
we can deduce the relation
\begin{eqnarray}
  \overset{(5)}{G}_{\mu\nu} = \overset{(4)}{G}_{\mu\nu} 
     + K_{\mu}{}^\lambda K_{\lambda\nu} - K K_{\mu\nu}
     -{1\over 2} g_{\mu\nu} \left( K^{\alpha\beta} K_{\alpha\beta} -K^2 \right)
     + C_{y\mu y\nu} + {3\over \ell^2} g_{\mu\nu} \ , 
\end{eqnarray}
where $C_{y\mu y\nu}$ is components of the 5-dimensional Weyl tensor
 and $\ell$ is the curvature scale set by the negative cosmological constant 
 in the bulk. 

Assuming the $Z_2$-symmetry, the junction condition leads to 
\begin{eqnarray}
 K^\mu{}_\nu - \delta^\mu_\nu K 
 = {\kappa^2 \over 2} \left( -\sigma \delta^\mu_\nu + T^\mu{}_\nu \right) \ , 
\end{eqnarray}
where $\kappa^2$, $\sigma$, and $T_{\mu\nu}$ are the coupling constant, 
the tension of the brane,  and the energy-momentum tensor for the matter
 on the brane. 

It is convenient to take the unit $\kappa^2 /\ell = 8\pi G=1$.
 By eliminating the extrinsic curvature in Eq.(3) using Eq.(4), we obtain
 the effective equation~\cite{ShiMaSa}
\begin{eqnarray}
  G_{\mu\nu} = T_{\mu\nu} + \ell^2 \pi_{\mu\nu} - E_{\mu\nu} \ ,
\end{eqnarray}
where
\begin{eqnarray}
 \pi_{\mu\nu} = -{1\over 4} T_{\mu\lambda} T^\lambda{}_\nu 
                + {1\over 12} T T_{\mu\nu} 
                +{1\over 8} g_{\mu\nu} \left( 
                T^{\alpha\beta} T_{\alpha\beta} -{1\over 3} T^2 \right)
\end{eqnarray}
and
\begin{eqnarray}
  E_{\mu\nu} = C_{y\mu y\nu} \Big|_{y=0}  \ .
\end{eqnarray}
Note that the projection of Weyl tensor $E_{\mu\nu}$ represents  the
 effect of the bulk geometry. 
Here, we have also assumed the relation $\kappa^2 \sigma = 6 / \ell $
 so that the effective cosmological constant vanish.

The geometrical approach is useful to classify possible corrections
 to the conventional Einstein equations. One defect of this approach
 is the fact that the projected Weyl tensor
 can not be determined without solving the equations in the bulk. 
 However, the traceless property
\begin{eqnarray}
     E^\mu{}_\mu =0
     \label{traceless}
\end{eqnarray}
is sufficient to determine the evolution of homogeneous universe.
 Indeed, the property (8) is the key to derive 
 the effective Friedman equation
\begin{eqnarray}
   H^2 = {1\over 3} \rho + \ell^2 \rho^2 + {C \over a^4} \ ,
\end{eqnarray}
where the constant of integration $C$ is referred to as the dark radiation.
 The effect of the bulk is encoded in this dark  radiation fluid. 
 Although its precise value of $C$ can not be determined without solving the
 bulk with the appropriate boundary condition, the dynamics of the
 homogeneous universe can be qualitatively understood.  
  For general spacetimes, however, this traceless condition is not sufficient
  to determine the evolution of the braneworld.


\section{Gradient Expansion of Effective Action}

 In this paper, for simplicity, we concentrate on the vacuum two-brane system. 
 Let us start with the 5-dimensional action  for this system
\begin{eqnarray}
  S [\gamma_{AB}, g_{\mu\nu}, h_{\mu\nu}] 
\end{eqnarray}
where $\gamma_{AB}$, $g_{\mu\nu}$ and $h_{\mu\nu}$ are the 5-dimensional
 bulk metric, the induced metric on the positive and the negative 
 tension branes, respectively. Here, $A,B$ and $\mu, \nu $ 
 label the 5-dimensional and 4-dimensional coordinates, respectively.   
 The variation with respect to $\gamma_{AB}$ gives the bulk Einstein equations
 and the variation with respect to $g_{\mu\nu}$ and $h_{\mu\nu}$ 
 yields the junction  conditions. 
 Now,  suppose to solve the bulk equations of motion and the 
 junction condition on the negative tension brane, 
 then formally we get the relation
\begin{eqnarray}
  \gamma_{AB} = \gamma_{AB}[g_{\mu\nu}] 
  \ , \quad h_{\mu\nu} = h_{\mu\nu} [g_{\mu\nu}] \ .
  \label{bulk-metric}
\end{eqnarray}
By substituting relations (\ref{bulk-metric})  into the original action,
 in principle, the 4-dimensional effective action can be obtained as
\begin{eqnarray}
 S_{\rm eff} 
 = S[\gamma_{AB} [g_{\mu\nu}] ,g_{\mu\nu}, h_{\mu\nu}[g_{\mu\nu}]] \ .
 \label{eff-action}
\end{eqnarray}
It should be stressed that the above effective action is nonlocal. 
 Moreover,  the above calculation  is not feasible in practice. 
 In reality, at low energy where the most of interesting phenomena occurs,
 we need not to follow the above general procedure. 
 In the following, we will give a method 
 to obtain the effective action without doing the above calculation.    

  The point is that the gradient expansion approach can be used  at low energy.
 At low energy, it is legitimate to assume that 
 the action can be expanded by the local terms with increasing
  orders of derivatives if one includes all of the massless modes~\cite{KS1}. 
 In the two-brane system, the relevant degrees of freedom
 are nothing but the metric  and  the radion  which represents the distance
 between two branes~\cite{GT}. 
  Hence, we assume the general local action constructed from the 
  metric $g_{\mu\nu}$ and the radion $\Psi $ as an ansatz.   
 Therefore,  we can write the action as 
\begin{eqnarray}
S_{\rm eff}
\!\!&=&\!\!
	{1\over 2} \int d^4 x \sqrt{-g} \left[ \Psi R - 2\Lambda (\Psi )
	-{\omega (\Psi) \over \Psi} \nabla^\mu \Psi \nabla_\mu \Psi \right]
        \nonumber\\
&&\!\!
	+\int d^4 x \sqrt{-g} \left[
	A(\Psi) \left( \nabla^\mu \Psi \nabla_\mu \Psi \right)^2
	+B(\Psi) \left( \Box \Psi \right)^2
	+C(\Psi)\nabla^\mu \Psi \nabla_\mu \Psi \Box \Psi
	\right. \nonumber \\
&&\!\!\!\!\left.
	+D(\Psi) R~\Box \Psi 
	+ E(\Psi) R \nabla^\mu \Psi \nabla_\mu \Psi
        + F(\Psi) R^{\mu\nu} \nabla_\mu \Psi \nabla_\nu \Psi 
        \right. \nonumber\\
&&\left.\!\!\!\!  
        + G(\Psi) R^2     
	+ H(\Psi) R^{\mu\nu} R_{\mu\nu} 
	+I(\Psi) R^{\mu\nu\lambda\rho} R_{\mu\nu\lambda\rho} 
	+\cdots  \right]  \ , 
	\label{setup}
\end{eqnarray}
where $\nabla_\mu$ denotes the covariant derivative with respect to the metric
$g_{\mu\nu}$ and $\Lambda, \omega , A, \cdots$ are  arbitrary
 coefficient functionals. 
 Here, we have listed up all of the possible local terms 
which have derivatives up to fourth-order. 
This can be regarded as  the generalization of the scalar-tensor theory
  including the higher derivative terms. 
 We have the freedom to redefine the scalar field $\Psi$. In fact, we have used
 this freedom to fix the functional form of the coefficient of the 
 Einstein-Hilbert term. 
 However, we can not determine other coefficient functionals without any 
 information about the bulk geometry.


\section{Strategy}

 In the gradient expansion approach, 
 we have introduced the radion explicitly. While
 the radion never appears in the geometric approach, instead  $E_{\mu\nu}$ 
 is induced as the effective energy-momentum tensor reflecting the
 effects of the bulk geometry.  
 Notice that the  property (\ref{traceless}) implies the conformal invariance 
 of this effective matter. Clearly, both approaches should agree to each other.
  Hence, the radion  must play a role of the conformally invariant matter 
 $E_{\mu\nu}$.   This requirement gives a stringent constraint on the
 action, more precisely,  the conformal symmetry (\ref{traceless}) determines
 radion dependent  coefficients  in the action (\ref{setup}). 

 Let us illustrate our method using the following action truncated at the second
 order derivatives:
\begin{eqnarray}
    S_{\rm eff} 
    ={1\over 2}  \int d^4 x \sqrt{-g} \left[ 
          \Psi R -2\Lambda (\Psi) - {\omega (\Psi) \over \Psi} 
          \nabla^\mu \Psi \nabla_\mu \Psi \right] \ ,
\end{eqnarray}
 which is nothing but the scalar-tensor theory with coupling function
 $\omega (\Psi)$ and the potential function $\Lambda (\Psi )$. 
 Note that this is the most general local action which contains 
 up to the second  order derivatives and has the general coordinate invariance.
 It should be stressed that the scalar-tensor theory is, in general,
  not related to the braneworld. However, we know a special type of 
 scalar-tensor theory corresponds to  the low energy 
 braneworld~\cite{KS1}. 
 Here, we will present a simple derivation of this known fact.  
 
 For the vacuum brane, i.e.  $T_{\mu\nu} \propto g_{\mu\nu}$, 
 we can put 
$T_{\mu\nu} + \ell^2 \pi_{\mu\nu} = - \lambda g_{\mu\nu}$. Hence, 
 the geometrical effective equation  reduces to 
\begin{eqnarray}
   G_{\mu\nu} =  - E_{\mu\nu} - \lambda g_{\mu\nu}   \ .
\end{eqnarray}

 First, we must find $E_{\mu\nu}$. 
The above action (14) gives the equations of motion for the metric as
\begin{eqnarray}
    G_{\mu\nu} &=& -{\Lambda \over \Psi} g_{\mu\nu}
                   + {1\over \Psi} \left( 
                 \nabla_\mu \nabla_\nu \Psi - g_{\mu\nu} \Box \Psi \right)
                 + {\omega \over \Psi^2} \left(
             \nabla_\mu \Psi \nabla_\nu \Psi  -{1\over 2} g_{\mu\nu}
             \nabla^\alpha \Psi \nabla_\alpha \Psi \right)  \ . 
\end{eqnarray}
The right hand side of this Eq.~(16) should be identified with 
$-E_{\mu\nu}-\lambda g_{\mu\nu}$.
 Hence, the  condition  $E^\mu{}_\mu =0$ becomes
\begin{eqnarray}
        \Box \Psi = - {\omega \over 3\Psi} 
        \nabla^\mu \Psi \nabla_\mu \Psi  
        - {4\over 3} \left( \Lambda  - \lambda \Psi \right)  \ .
\end{eqnarray}
This is the equation for the radion $\Psi$. However, we also
have the equation for $\Psi$ from the action (14) as 
\begin{eqnarray}
    \Box \Psi = \left( {1\over 2\Psi} - { \omega' \over 2\omega} \right)
             \nabla^\alpha \Psi \nabla_\alpha \Psi  
             - {\Psi \over 2\omega} R + {\Psi \over \omega} \Lambda'    \ ,
\end{eqnarray}
where the prime denotes the derivative with respect to $\Psi$. 
In order for these two Eqs.~(17) and (18) to be compatible, $\Lambda$ and
  $\omega$ must satisfy 
\begin{eqnarray}
&&-{\omega \over 3 \Psi} = {1\over 2\Psi} - { \omega' \over 2\omega}  \ ,  \\
&&{4\over 3} \left( \Lambda - \lambda \Psi \right) =
            {\Psi \over \omega} \left( 2\lambda - \Lambda' \right)  \ ,
\end{eqnarray}
where we used  $R= 4\lambda$ which comes from the trace part of Eq.~(15).
 Eqs.~(19) and (20) can be integrated as  
\begin{eqnarray}
   \Lambda (\Psi) = \lambda + \lambda \beta \left( 1-\Psi \right)^2   \ , \quad
  \omega (\Psi ) = {3\over 2} {\Psi \over 1-\Psi}  \ ,  
\end{eqnarray}
where the constant of integration $\beta$ represents the ratio
 of the cosmological constant on the negative tension brane to that on 
 the positive tension brane. 
 Here, one of constants of integration is absorbed by rescaling of $\Psi$.
 In doing so, we have assumed the constant of integration is positive.
 We can also describe the negative tension brane if we take the 
 negative signature.

Thus, we get the effective action 
\begin{eqnarray}
S_{\rm eff}
	&=&\int d^4 x \sqrt{-g} \left[ {1\over 2} \Psi R 
	-{3 \over 4( 1-\Psi )} \nabla^\mu \Psi \nabla_\mu \Psi 
    - \lambda - \lambda \beta (1-\Psi)^2 \right] \ .
\end{eqnarray}
Surprisingly, this completely agrees with the previous 
result~\cite{KS1}. Our simple symmetry 
principle $E^\mu{}_\mu =0$ has determined the action completely. 
 
 As we have shown in \cite{KSS}, if $\beta <-1$
 there exists a static deSitter two-brane solution
 which turns out to be unstable. In particular,  
 two inflating branes  can collide at $\Psi = 0$.


\section{KK corrections and Implications}

  Let us apply the procedure in the previous section
 to the higher order case. 
 From the linear analysis, the action in the previous section is known 
 to come from zero modes. Hence, one can expect the other coefficients 
 in the action (\ref{setup}) represent the effects of KK-modes. 
 
  Now we impose the conformal symmetry $E^\mu{}_\mu =0$
   on the fourth order derivative terms
  in the action (\ref{setup}) as we did in the  previous section. 
  Starting from the action (\ref{setup}), one can read off the equation for 
  the metric from which $E_{\mu\nu}$ can be identified. 
  From the compatibility between the equations of motion for $\Psi$
  and the equation $E^\mu{}_\mu =0$ determines the coefficient functionals
  in the action (\ref{setup}). 
 The compatibility condition between $E^\mu{}_\mu =0$ and the equation for the
 radion $\Psi$ leads to
\begin{eqnarray}
  &&  (1-\Psi) (C'' -3A') = C' + 3E''+{3\over 2} F'' \\
  &&  (1-\Psi) (2B'' -4A) = 2B' +C + 3D'' + 3E'+{5\over 2} F' \qquad \\
  &&  (1-\Psi) (4C' -8A) = 2C + 12 E' + 5F' \\
  &&  (1-\Psi) (3B'-2C) = 2B + 3 D' + F   \\ 
  &&  4(1-\Psi) B' = 2B + 6 D' + 6E + 3F \\
  &&  (1-\Psi) C =  3E +F   \\
  && 2 (1-\Psi) B = 3 D \\
  &&  (1-\Psi) (2C - F') = 6E + 2F + 2H'' + 4I'' \\
  &&  (1-\Psi) F = - H' - 2 I' \\
  && (1-\Psi ) (D'' - E' ) = D' + 6 G'' + H''  \\
  &&  2 (1-\Psi ) (D' -E ) = D + 6 G' + H'   \\
  &&  G' = H' = I' = 0  \ .
\end{eqnarray}
These equations seem to be over constrained. Nevertheless, one can find
 solutions consistently. 
From Eqs.~(31) and (34), we see $F=0$. 
Eqs.~(28) and (29) can be solved as
\begin{eqnarray}
 E = {1\over 3} (1-\Psi ) C \ , \quad 
   D =  {2\over 3} (1-\Psi ) B \ . 
\end{eqnarray}
Substituting these results into Eqs.~(23) -(27) and Eqs.~(32) and (33), we have
\begin{eqnarray}
  && 3 (1-\Psi ) A' = C'  \ , \\ 
  && (1-\Psi ) (C' + 4A ) = 2B' \ ,\\
  && 4(1-\Psi ) A = C  \ , \\
  && B'-2C = 0   \ , \\
  &&  (1-\Psi ) C = B \ , \\
  &&  2B = (1-\Psi ) \left[ 
      6B' - 2(1-\Psi ) B'' -C + (1-\Psi ) C' \right] \ , \qquad \\
  && 3B = (1- \Psi ) (2B' -C)   \ .
\end{eqnarray}
Combining  Eqs.~(39) and (40), we obtain
\begin{eqnarray}
      B = {\ell^2 \over (1-\Psi )^2 }  \ , 
\end{eqnarray}
where $\ell^2$ is the constant of integration representing the curvature
 scale of the bulk.  
 Eqs.~(40) and (38) give 
\begin{eqnarray}
    C= {\ell^2 \over (1-\Psi )^3} \ , \quad 
    A= {1\over 4}{\ell^2 \over (1-\Psi)^4} \ .
\end{eqnarray}
The rest of Eqs.~(36), (37), (41) and (42) are identically satisfied. 
 The coefficients $G, H$ and $I$ must be constants $g,h$ and $i$.  
 Because of the existence of the Gauss-Bonnet topological term, we can put $i=0$
 without losing the generality.

Thus, we find the 4-dimensional effective action with KK corrections as
\begin{eqnarray}
S_{\rm eff} 
  &=& \int d^4 x \sqrt{-g} \left[ {1\over 2} \Psi R 
     - {3 \over 4( 1-\Psi )} \nabla^\mu \Psi \nabla_\mu \Psi 
    - \lambda - \lambda \beta (1-\Psi)^2    \right]
              \nonumber\\
 &&    + \ell^2 \int d^4 x \sqrt{-g} \left[
     {1 \over 4 (1-\Psi)^4} \left( \nabla^\mu \Psi \nabla_\mu \Psi \right)^2
      + {1\over (1-\Psi)^2} \left( \Box \Psi \right)^2 
      + {1\over (1-\Psi)^3} \nabla^\mu \Psi \nabla_\mu \Psi\Box \Psi
      \right. \nonumber \\
 && \left.   + {2\over 3(1-\Psi)} R \Box \Psi 
      + {1\over 3(1-\Psi)^2}  R \nabla^\mu \Psi \nabla_\mu \Psi 
      + g R^2     + h R^{\mu\nu} R_{\mu\nu}  \right] \ ,
 \label{action}
\end{eqnarray}
where constants $g$ and $h$ can be interpreted as 
  the effects of the bulk gravitational waves.

From the point of view of the geometrical equation
\begin{eqnarray}
 \delta G_{\mu\nu} = \delta T_{\mu\nu} + \ell^2 \delta \pi_{\mu\nu}
   - \delta E_{\mu\nu}  \ ,
\end{eqnarray}
the key to understand the evolution of the cosmological perturbations 
 is to understand $E_{\mu\nu}$.
 In the conformal Newtonian gauge
\begin{eqnarray}
   ds^2 = - (1+2\Psi ) dt^2 + a^2 (t) (1+ 2\Phi) \delta_{ij} dx^i dx^j ,
\end{eqnarray}
the statement can be more explicit, namely,  the effective anisotropic 
stress $\Psi -\Phi$ should be known.  
Once it is determined, the anisotropy of the cosmic background radiation 
\begin{eqnarray}
{\delta T \over T} = \zeta + \Psi -\Phi
\end{eqnarray}
can be calculated. Here, $\zeta$ denotes the curvature perturbation.  
 Starting our action (23), we can determine $E_{\mu\nu}$ which
  include Kaluza-Klein effect. 
  Thus, this anisotropic stress can be  determined.


\section{Conclusion}

 We have combined the geometrical approach and the low energy
 gradient expansion method to derive the 4-dimensional 
 effective action. Given our effective action, the temperature
 anisotropy of the cosmic background radiation can be computed
 without solving the bulk  equations of motion explicitly.
 Since the method for completing this part is standard, we have not 
 spelled out details here~\cite{soda}. 
 We have just mentioned the importance of
 our result in conjunction with the anisotropy of temperature
 of the cosmic background radiation. 
 
\begin{acknowledgements}
This work was supported in part by  Grant-in-Aid for  Scientific
Research Fund of the Ministry of Education, Science and Culture of Japan 
 No. 155476 (SK) and  No.14540258 (JS) and also
  by a Grant-in-Aid for the 21st Century COE ``Center for
  Diversity and Universality in Physics".  
\end{acknowledgements}


\begin{thebibliography}{99}

\bibitem{RS1}
L.~Randall and R.~Sundrum,
Phys.\ Rev.\ Lett.\  {\bf 83}, 3370 (1999).


\bibitem{KS1}
S.~Kanno and J.~Soda,
Phys.\ Rev.\ D {\bf 66}, 083506 (2002);
S.~Kanno and J.~Soda,
Phys.\ Rev.\ D {\bf 66}, 043526 (2002);
S.~Kanno and J.~Soda,
Gen.\ Rel.\ Grav.\  {\bf 36}, 689 (2004);
S.~Kanno and J.~Soda,
arXiv:hep-th/0407184.

\bibitem{Sugumi}
S.~Kanno and J.~Soda, Phys. Lett. B588, 203 (2004). 


\bibitem{ShiMaSa}
T.~Shiromizu, K.~Maeda and M.~Sasaki,
Phys.\ Rev.\ D {\bf 62}, 024012 (2000).

\bibitem{GT}
J.~Garriga and T.~Tanaka,
Phys.\ Rev.\ Lett.\  {\bf 84}, 2778 (2000).


 
\bibitem{KSS}
S.~Kanno, M.~Sasaki and J.~Soda,
Prog. Theor. Phys. {\bf 109}, 357 (2003).



\bibitem{soda}
D.~Langlois,
Phys.\ Rev.\ Lett.\  {\bf 86}, 2212 (2001);
H.~Kodama, A.~Ishibashi and O.~Seto,
Phys.\ Rev.\ D {\bf 62}, 064022 (2000);
C.~van de Bruck, M.~Dorca, R.~H.~Brandenberger and A.~Lukas,
Phys.\ Rev.\ D {\bf 62}, 123515 (2000);
K.~Koyama and J.~Soda,
Phys.\ Rev.\ D {\bf 62}, 123502 (2000);
D.~Langlois, R.~Maartens, M.~Sasaki and D.~Wands,
Phys.\ Rev.\ D {\bf 63}, 084009 (2001);
K.~Koyama and J.~Soda,
Phys.\ Rev.\ D {\bf 65}, 023514 (2002).




\end{thebibliography}

\end{document}